\documentclass[prd,twocolumn,floatfix,reprint,aps,superscriptaddress]{revtex4}

\usepackage{amsfonts}
\usepackage{mathrsfs}
\usepackage{amsmath}
\usepackage{color}
\usepackage{graphicx}
\usepackage{bm}
\usepackage{amssymb}
\usepackage{xspace}
\usepackage{epstopdf}
\usepackage{dcolumn}
\usepackage{longtable}
\usepackage{multirow}
\usepackage[colorlinks=true, letterpaper=true, pdfstartview=FitV, linkcolor=blue, citecolor=blue, urlcolor=blue]{hyperref}

\begin{document}

\title{Monolayer Mg$_{2}$C: Negative Poisson's ratio and unconventional 2D emergent fermions}

\author{Shan-Shan Wang}
\affiliation{Research Laboratory for Quantum Materials, Singapore University of Technology and Design, Singapore 487372, Singapore}

\author{Ying Liu}
\affiliation{Research Laboratory for Quantum Materials, Singapore University of Technology and Design, Singapore 487372, Singapore}

\author{Zhi-Ming Yu}
\affiliation{Research Laboratory for Quantum Materials, Singapore University of Technology and Design, Singapore 487372, Singapore}

\author{Xian-Lei Sheng}
\affiliation{Department of Applied Physics, Key Laboratory of Micro-nano Measurement-Manipulation and Physics (Ministry of Education), Beihang University, Beijing 100191, China}

\author{Liyan Zhu}
\email{lyzhu@hytc.edu.cn}
\affiliation{School of Physics and Electronic $\&$ Electrical Engineering, and Jiangsu Key Laboratory of Modern Measurement Technology and Intelligent Systems, Huaiyin Normal University, Huai'an, Jiangsu 223300, China}

\author{Shan Guan}
\email{physguan@gmail.com}
\affiliation{Research Laboratory for Quantum Materials, Singapore University of Technology and Design, Singapore 487372, Singapore}
\affiliation{Beijing Key Laboratory of Nanophotonics and Ultrafine Optoelectronic Systems, School of Physics,
Beijing Institute of Technology, Beijing 100081, China}

\author{Shengyuan A. Yang}
\affiliation{Research Laboratory for Quantum Materials, Singapore University of Technology and Design, Singapore 487372, Singapore}

\begin{abstract}
    Novel two-dimensional (2D) emergent fermions and negative Poisson's ratio in 2D materials are fascinating subjects of research. Here, based on first-principles calculations and theoretical analysis, we predict that the hexacoordinated Mg$_{2}$C monolayer hosts both exotic properties.
    We analyze its phonon spectrum, reveal the Raman active modes, and show that it has small in-plane stiffness constants. Particularly, under the tensile strain in the zigzag direction, the Mg$_{2}$C monolayer shows an intrinsic negative Poisson's ratio $\sim -0.023$, stemming from its unique puckered hinge structure.
    The material is metallic at its equilibrium state. A moderate biaxial strain can induce a metal-semimetal-semiconductor phase transition, during which several novel types of 2D fermions emerge, including the anisotropic Dirac fermions around 12 tilted Dirac points in the metallic phase, the $2$D double Weyl fermions in the semimetal phase where the conduction and valence bands touch quadratically at a single Fermi point, and the 2D pseudospin-1 fermions at the critical point of the semimetal-semiconductor phase transition where three bands cross at a single point on the Fermi level. In addition, uniaxial strains along the high-symmetry directions break the three-fold rotational symmetry and reduce the number of Dirac points. Interestingly, it also generates 2D type-II Dirac points. We construct effective models to characterize the properties of these novel fermions. Our result reveals Mg$_{2}$C monolayer as an intriguing platform for the study of novel 2D fermions, and also suggests its great potential for nanoscale device applications.
\end{abstract}


\maketitle

\section{Introduction}

Two-dimensional (2D) materials have been attracting tremendous interest~\cite{novoselov2005two,manzeli20172d,bhimanapati2015recent} since the discovery of graphene~\cite{novoselov2004}. The many extraordinary properties of graphene can be attributed to its unique electronic band structure~\cite{neto2009}: the conduction and valence bands linearly touch at the two corner points of the Brillouin zone (BZ), around which the quasi-particles can be described by 2D massless Dirac fermions. Inspired by graphene, much effort has been devoted to exploring other 2D material with band crossings points. Many candidates have been proposed to host linear Dirac points~\cite{wang2015rare}, such as the group-IV monolayers like silicene~\cite{cahangirov2009two,liu2011quantum} and germanene~\cite{cahangirov2009two,liu2015multiple}, 2D carbon and boron allotropes~\cite{xu2014two,zhou2014semimetallic,ma2016graphene,jiao2016two,feng2017dirac}, group-V monolayers~\cite{lu2016multiple,kim2015observation}, and $5d$ transition metal trichlorides~\cite{sheng2017monolayer}. Beyond the 2D Dirac fermion, other types of 2D fermions may also exist. For example, 2D double Weyl and 2D pseudospin-$1$ fermions were proposed in the blue phosphorene oxide monolayer~\cite{zhu2016blue}. 2D spin-orbit Dirac fermions were predicted in monolayer HfGeTe family materials~\cite{guan2017two}. In addition, 2D nodal-loop fermions have been proposed in monolayer metal-group VI compounds~\cite{jin2017prediction} and monolayer $X_3$SiTe$_6$ ($X=$Ta, Nb)~\cite{li2018nonsymmorphic}.
So far, the discovered 2D materials with novel emergent fermions are still limited. Therefore, it is much desired to explore new 2D materials which may host new types of fermionic excitations.

Meanwhile, the transition metal carbides, which belong to the so-called MXene family, have emerged as a new class of 2D materials~\cite{naguib2012two,anasori20172d}. Efficient methods for synthesizing quite a number of MXene materials (such as V$_2$C, Ti$_{2}$C, Nb$_{2}$C, Mo$_{2}$C, and \emph{etc.}) have been developed~\cite{halim2016synthesis}. Besides the transition metals, the main-group metal elements can also form the MXene structure. For instance, both Be$_{2}$C and Mg$_{2}$C monolayers have been proposed~\cite{li2014be2c,naseri2017magnesium,meng2017metal}, which possess quasi-planar structure with small buckling, analogous to the MXenes. Particularly, the MXene structure for Mg$_{2}$C is predicted as the global minimum from the particle swarm optimization method, and is found to be stable at temperatures as high as 900 K~\cite{meng2017metal}.  It is also worth noting that another 2D hexacoordinated structure Cu$_{2}$Si was proposed to host nodal lines, which was recently verified in experiment~\cite{feng2017experimental,yang2015two}. In addition, some of the MXenes also exhibit nontrivial band topology~\cite{fashandi2015dirac,weng2015large,si2016large}. Thus, it is reasonable to assume that novel 2D emergent fermions may arise in some of these new 2D materials.

In this work, based on first-principle calculations and theoretical analysis, we reveal that the hexacoordinate-arranged Mg$_{2}$C monolayer actually provides an intriguing playground for studying multiple types of fermionic excitations which are controllable by strain. We analyze the phonon spectrum for the Mg$_{2}$C monolayer, reveal its Raman active modes, and show that it has small in-plane stiffness constants.
Particularly, due to the puckered hinge structure, the material possesses a negative intrinsic Poisson's ratio under applied strain along the zigzag direction. At the equilibrium state, the Mg$_{2}$C monolayer demonstrates a quadratic band touching at the $\Gamma$ point, with 12 tilted Dirac points on the high-symmetry paths in the BZ close to Fermi level. An applied biaxial strain can drive a metal-semimetal-semiconductor quantum phase transition. In the semimetal phase, the quadratic band touching at $\Gamma$ features $2$D double Weyl fermions. And at the critical point of the semimetal-semiconductor phase transition, three low-energy bands touch at a single point at the Fermi level, resulting in 2D pseudospin-$1$ fermions. These novel quasi-particles possess unique properties such as universal optical absorbance, super Klein tunneling, and super collimation. In addition, we find that the uniaxial strain along high-symmetry directions break the three-fold rotational symmetry but preserves a reduced number of Dirac points, including 2D type-II Dirac point.
We construct effective models to characterize these novel emergent fermions. Our findings provide a promising material platform for exploring the
fundamental physics associated with unconventional emergent fermions in two dimensions. The revealed excellent mechanical properties and efficient strain-tuning of phase transitions suggest the material's great potential in
nanoscale device applications.

\section{Computational Methods}
Our first-principles calculations were based on the density functional theory (DFT), as implemented in the Vienna \emph{ab initio} simulation package (VASP)~\cite{kresse1993ab,kresse1996g}. The interactions between electrons and ions were modeled by the projector augmented wave method~\cite{blochl1994pe}. The generalized gradient approximation (GGA) parameterized by the Perdew, Burke, and Ernzerhof (PBE) was adopted to describe the exchange and correlation interactions~\cite{perdew1996generalized}. A vacuum space of $20$ {\AA} was used to avoid the artificial interactions between periodic images. The energy cutoff was set to be $520$ eV for the plane-wave basis. The Monkhorst-Pack $k$-mesh with size $45\times45\times1$ was adopted for the Brillouin zone sampling. The lattice parameters and the ionic positions were fully optimized until the residual force on each atom was less than $0.01$ eV/{\AA}. And the energy convergence criterion was set to be $10^{-7}$ eV.

\section{Crystal structure and Phonon spectrum}

\begin{figure}[t!]
\includegraphics[width=7.8cm]{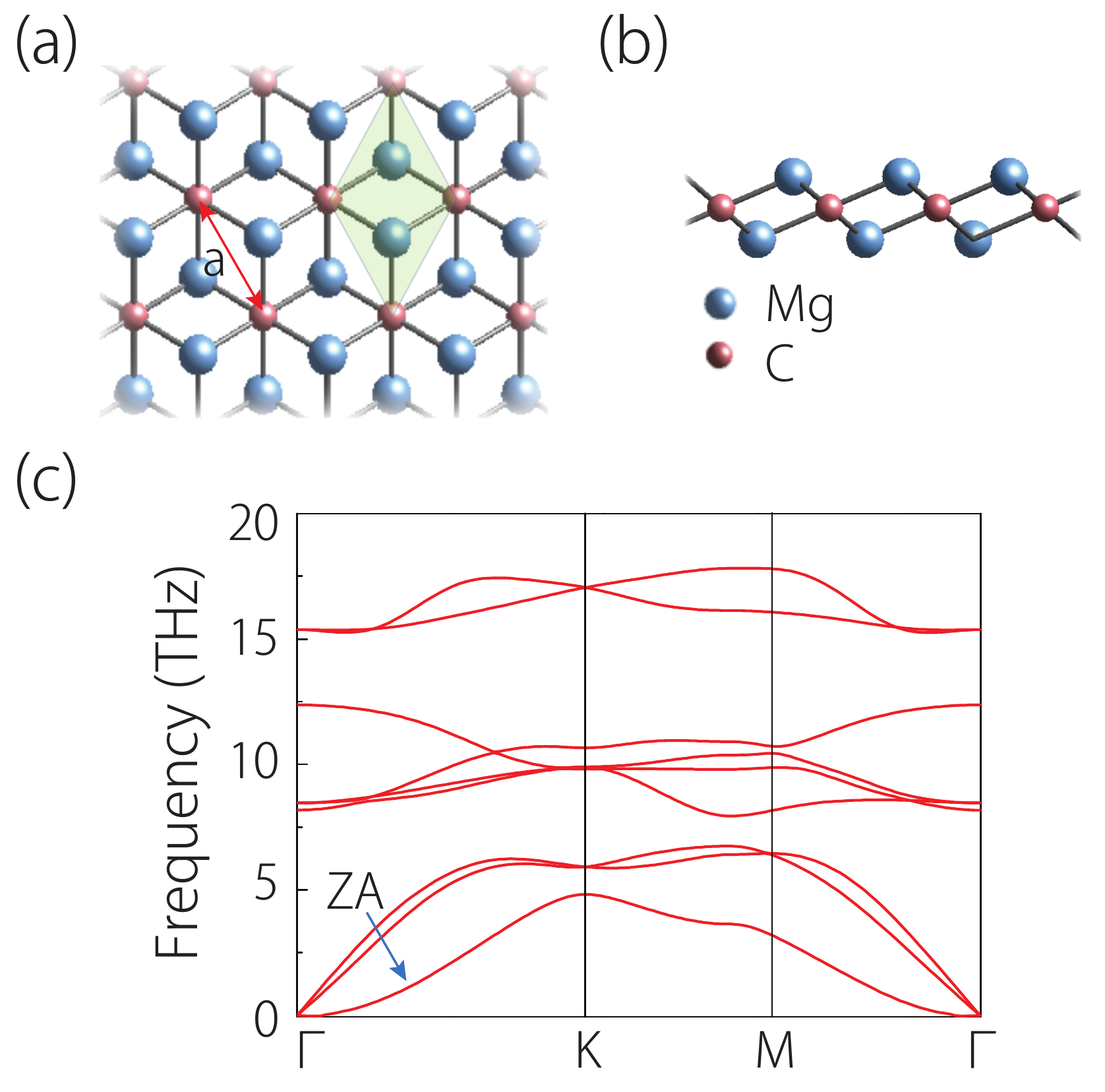}
\caption{(a) Top and (b) side view of the lattice structure for 2D Mg$_{2}$C monolayer. The green shaded region in (a) indicates the primitive unit cell. $a$ is the lattice parameter. (c) Phono spectrum for the Mg$_{2}$C monolayer. A $6\times6$ supercell is used in the calculation.
}
\label{fig1}
\end{figure}

The lattice structure for the Mg$_{2}$C monolayer is shown in Fig.~\ref{fig1}(a,b). It possesses three hexagonal atomic layers: a basal plane of carbon sandwiched by two layers of Mg atoms with an $ABC$-type stacking. Each C atom is bonded to six Mg atoms. The lattice has a point (space) group symmetry of $D_{3d}$ $(P\bar{3}m1)$, with two Mg and one C atoms in a primitive unit cell. The equilibrium lattice constant is $3.549$ {\AA} for the fully relaxed structure, in good agreement with the previous work~\cite{meng2017metal}. The equilibrium lengths of the bonds Mg-Mg and Mg-C are $2.712$ {\AA} and $2.233$ {\AA}, respectively. Compared with previous reported Be$_{2}$C monolayer~\cite{li2014be2c}, the Mg$_{2}$C monolayer share the similar structure, only with larger buckling, i.e., larger separation between the atomic layers.

The dynamical stability of the Mg$_{2}$C structure can be inferred from the phonon spectrum. As shown in Fig.~\ref{fig1}(c), one observes that there is no imaginary phonon frequency throughout the BZ, indicating that the Mg$_{2}$C monolayer is dynamically stable. As approaching the $\Gamma$ point, while linear dispersions are observed for two acoustic branches, the frequency of the out-of-plane (ZA) acoustic mode exhibits a quadratic dependence on the wave vector, which is a characteristic feature of layered materials~\cite{zabel2001phonons,zhu2014coexistence,carrete2016physically}. The sound speed of the longitudinal acoustic phonons ($\sim 5.3$ km/s) for Mg$_{2}$C is smaller than that of the pristine blue phosphorene ($\sim 8.1$ km/s)~\cite{zhu2014semiconducting}, MoS$_2$ ($\sim 6.5$ km/s)~\cite{zhu2015thermal}, as well as graphene ($\sim 21.2$ km/s)~\cite{kaasbjerg2012unraveling}. The smaller sound speed indicates that the in-plane stiffness of Mg$_{2}$C should be relatively small, as we show in the following.

Since the Mg$_2$C monolayer has a point group symmetry of $D_{3d}$, the optical phonon modes can be decomposed into
\begin{equation}
\Gamma_\text{optical} = E_g(R) + A_{1g}(R) + A_{2u} + E_u,
\end{equation}
where the symbol $R$ in the parentheses indicates that two modes are Raman active, namely, the intralayer shear ($E_g$) and breathing ($A_{1g}$) modes. These Raman active modes are sensitive to the externally applied strain~\cite{meng2017metal}, hence can be used as a tool to characterize the strain distribution in the monolayer.

\section{Elastic property and negative Poisson's ratio}

\begin{figure}[t!]
\includegraphics[width=8.4cm]{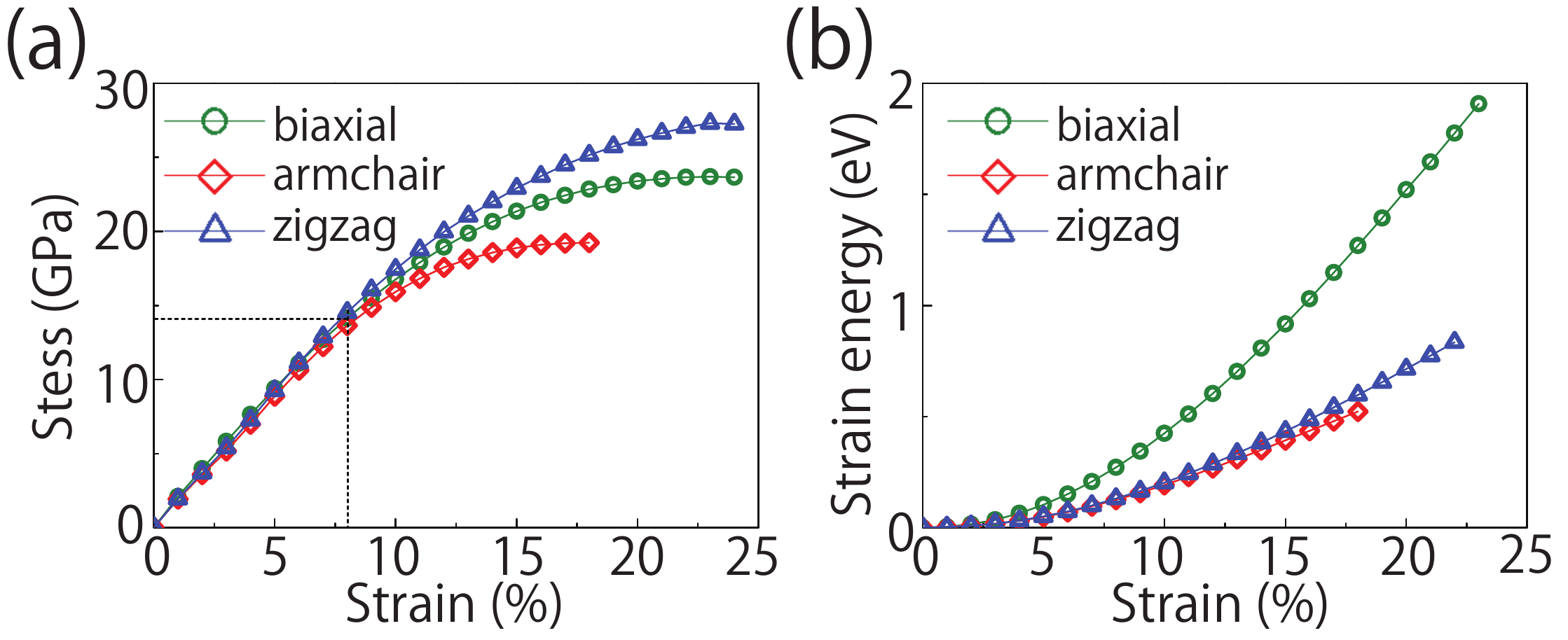}
\caption{(a) Strain-stress relation for the Mg$_{2}$C monolayer with different types of strain. The dashed line indicates the linear elastic regime. (b) Strain energy ($E_{s}$) as a function of applied strain. }
\label{figss}
\end{figure}

To investigate the elastic property of monolayer Mg$_{2}$C, particularly its susceptibility to strain, we calculate the strain-stress curves. The results are shown in Fig.~\ref{figss}(a). It is observed that monolayer Mg$_{2}$C remains within the linear elastic regime until about $8\%$ biaxial strain, and the material can sustain a biaxial strain up to $20\%$, while the critical uniaxial strain can be $\sim 18\%$ for the armchair direction and $\sim 22\%$ for the zigzag direction. These values are comparable to other 2D materials such as graphene, MoS$_{2}$, and Ti$_{2}$C~\cite{kim2009large,castellanos2012elastic,chakraborty2017manipulating}. For small deformations, the elastic property of $2$D material is usually characterized by the in-plane stiffness constant, defined as
\begin{equation}
C=\frac{1}{S_{0}}\frac{\partial^{2}E_{s}}{\partial\varepsilon^{2}},
\end{equation}
where $S_{0}$ is the equilibrium area of the unit cell, $E_{s}$ is the strain energy (i.e., the energy difference between the strained and unstrained systems), and $\varepsilon$ is the in-plane uniaxial strain. The calculated stain-energy curve is plotted in Fig.~\ref{figss}(b), from which the typical quadratic dependence on strain can be observed at small deformations. The obtained stiffness constants are $59.62$ N/m and $56.78$ N/m for strains along the armchair and the zigzag directions, respectively. Such stiffness constants are smaller than other typical 2D materials such as graphene ($\sim 340\pm 40$ N/m)~\cite{lee2008measurement}, MoS$_{2}$ ($\sim 140$ N/m)~\cite{peng2013outstanding}, and BN ($\sim 267$ N/m)~\cite{topsakal2010response}, which indicates that the Mg$_{2}$C monolayer is softer. This great flexibility will facilitate the strain engineering of its physical properties.

\begin{figure}[t!]
\includegraphics[width=8.5cm]{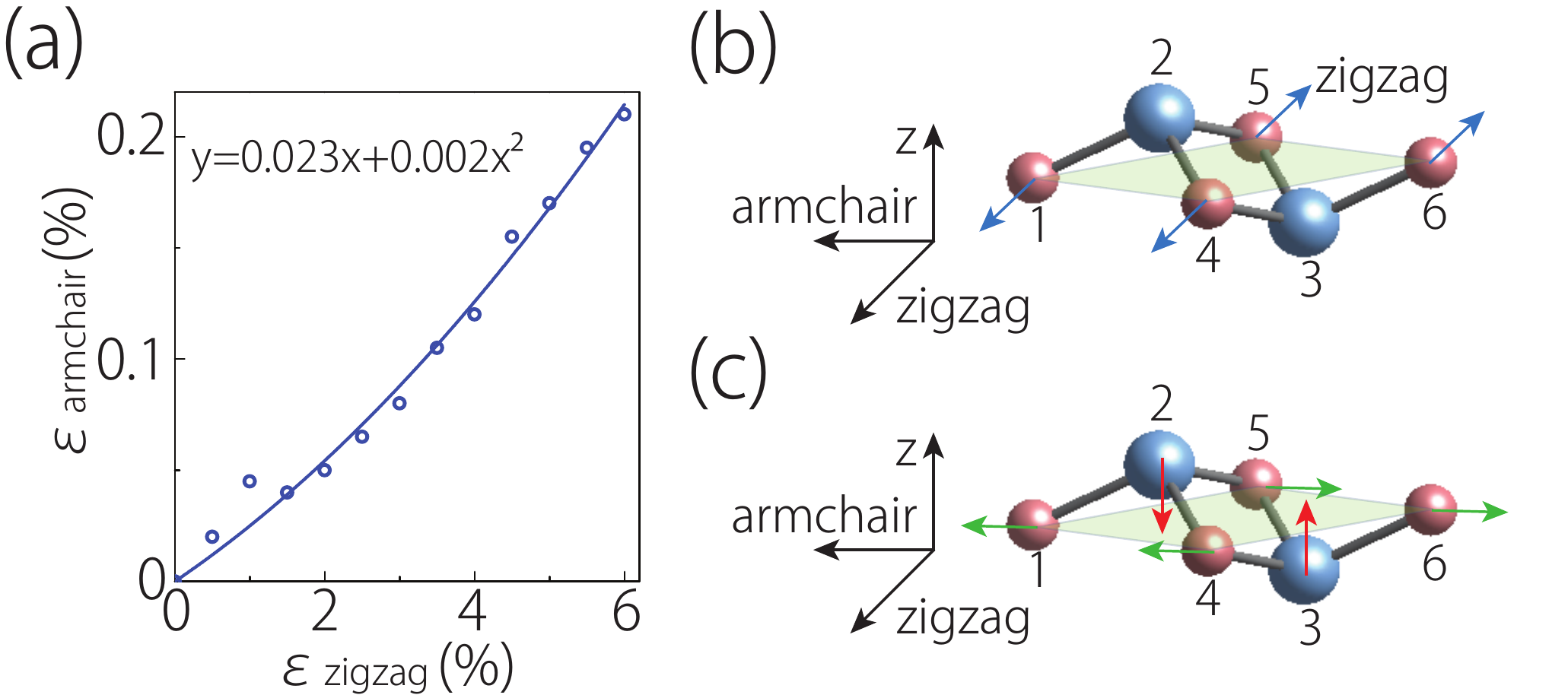}
\caption{(a) Responsive strain in the armchair direction induced by applied strain in the zigzag direction. (b) Illustration for Mg$_2$C stretched along the zigzag direction. Atoms are moved in the direction of the attached arrows. (c) To accommodate the tension in the zigzag direction, the Mg atoms move inward, towards the C plane. And consequently, the structure expand in the armchair direction, resulting in a negative Poisson's ratio.
In (b,c), the shaded region marks the C plane.}
\label{figNPR}
\end{figure}

The Poisson's ratio ($\nu=-\varepsilon_\text{transverse}/\varepsilon_\text{axial}$) characterizes the material's response to uniaxial strains. It is positive for most materials, which means that a material typically constricts in the transverse direction when it is stretched along the longitudinal direction~\cite{evans2000auxetic,lakes1993advances}. Interestingly, we find that the Mg$_{2}$C monolayer exhibits a negative Poisson's ratio when the applied strain is in the zigzag direction.
Fig.~\ref{figNPR}(a) shows the responsive strain in the armchair direction versus the applied strain in zigzag direction. Evidently, the monolayer tends to expand when subjected to a tensile strain in the zigzag direction. The value of the corresponding Poisson's ratio is $-0.023$, which is comparable with the previously found negative Poisson's ratio for black phosphorene ($\sim-0.022$)~\cite{jiang2014negative}.

To unveil the underlying mechanism, in Fig.~\ref{figNPR}(b) and \ref{figNPR}(c), we illustrate the movement of atoms when strain is applied. Under a tensile strain applied along the zigzag directions, the Carbon atoms will displace along the zigzag direction [indicated by the blue arrows in Fig.~\ref{figNPR}(b)]. To bear with the elongation in zigzag direction, the monolayer is compressed in the $z$ direction, which means that Mg atoms will move inward along $z$ as shown by red arrows in Fig.~\ref{figNPR}(c). The inward movement of Mg atoms will increase the bond angels like $\theta_{124}$ and $\theta_{125}$, because the bond lengths are kept more or less unchanged due to the strong inter-atomic forces. This mechanism compensates and surpasses the natural tendency of compression in the transverse direction.
As a consequence, it leads to an overall negative Poisson's ratio [see Fig.~\ref{figNPR}(c)]. We also mention that for strains applied along the armchair direction, the natural compression in the zigzag direction still dominates, giving a positive Poisson's ratio $\sim 0.03$.

\section{Strain-tunable emergent fermions}

Having established that the Mg$_2$C monolayer has excellent elastic properties and is susceptible to strain, we turn to investigate its electronic properties and how the properties can be controlled by strain. In the following, we show that a moderate strain can drive quantum phase transitions in the Mg$_2$C monolayer, and in the process there emerge multiple types of 2D unconventional emergent fermions.

\begin{figure}[t!]
\includegraphics[width=8.2cm]{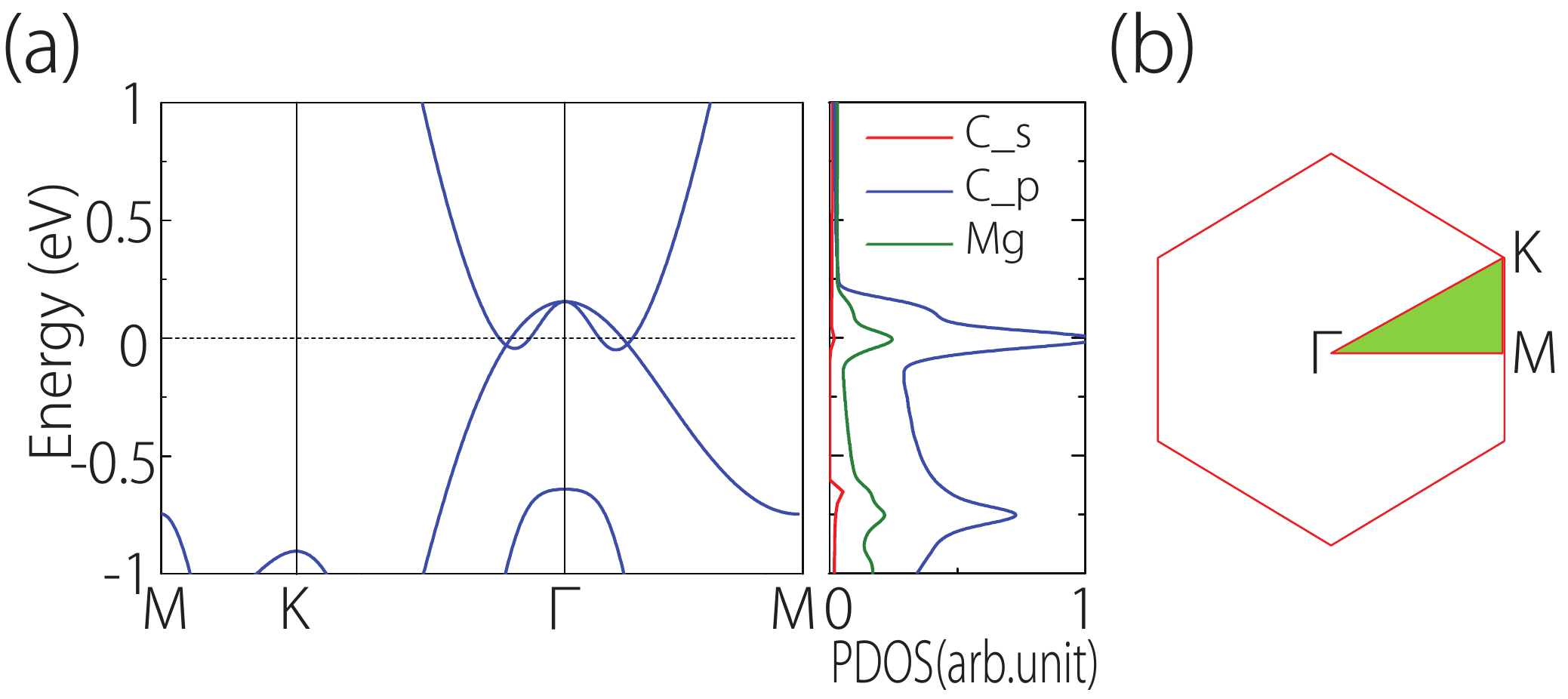}
\caption{(a) Left panel shows the electronic band structure for the Mg$_{2}$C monolayer at equilibrium state. Right panel shows the projected density of states (PDOS). (b) Brillouin zone with the high-symmetry points labeled.}
\label{figBand}
\end{figure}

\subsection{Quantum phase transition driven by biaxial strain}

\begin{figure}[t!]
\includegraphics[width=8cm]{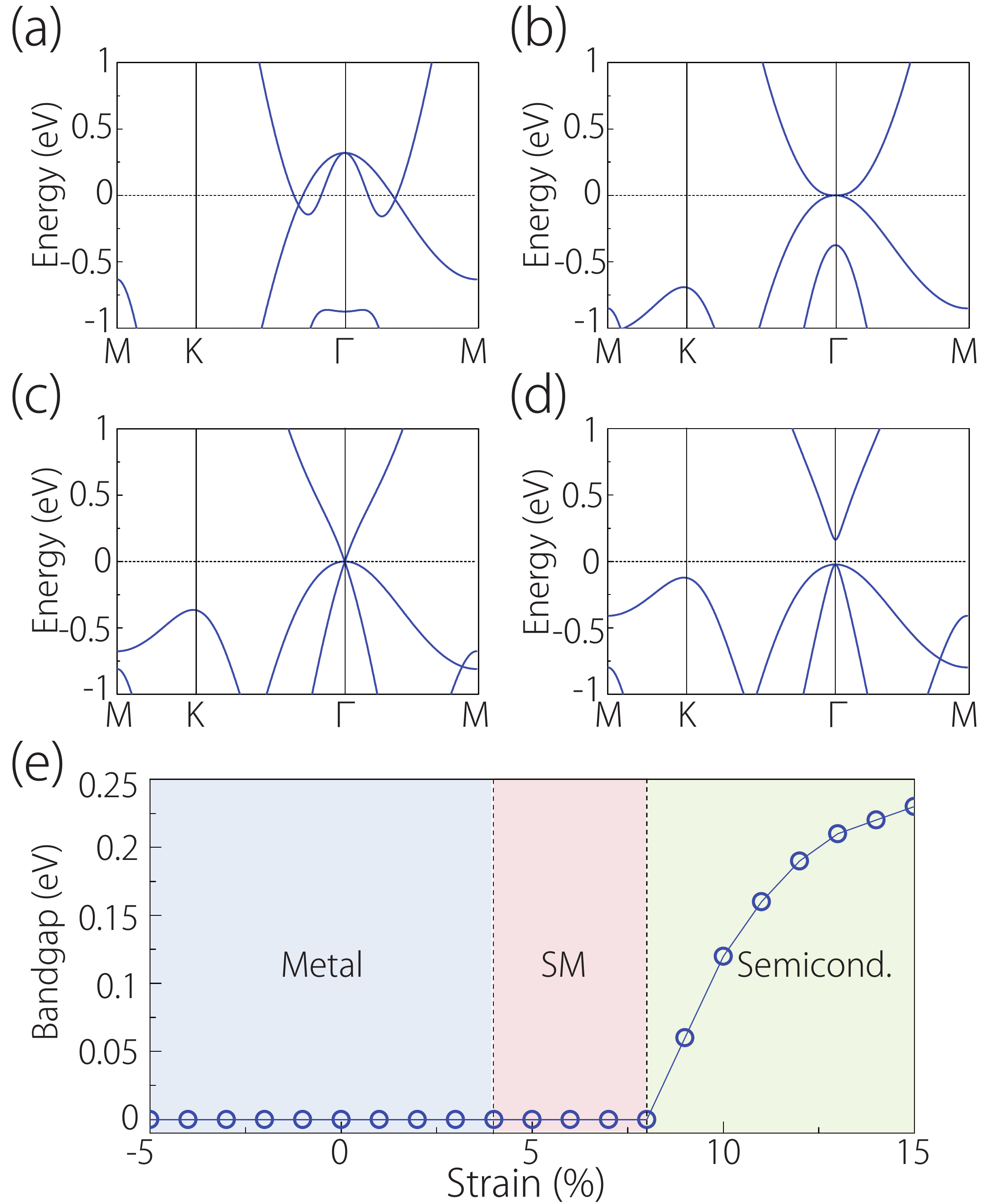}
\caption{(a) Electronic band structures for the Mg$_{2}$C monolayer under biaxial strain of (a) $-4\%$, (b) $+4\%$, (c) $+8\%$, and (d) $+11\%$.
(e) Phase diagram under biaxial strain. The data points indicate the value of bandgap.}
\label{figQPT}
\end{figure}

The electronic band structure at equilibrium (unstrained) state is shown in Fig.~\ref{figBand}. The material demonstrates a metallic phase. There are two bands crossing the Fermi level. From the projected density of states (PDOS), the low-energy states are mainly contributed by the Carbon $p_x$ and $p_y$ orbitals. As Carbon is a very light element, this also indicates that the effect of spin-orbit coupling (SOC) on the low-energy band structure will be negligibly small, which is verified by our explicit calculation (not shown). Hence, SOC is neglected in the following discussion.

For the two low-energy bands in Fig.~\ref{figBand}(a), one takes a hole-like parabolic dispersion, whereas the other one has a Mexican-hat-like shape. One notes that the two bands touch in a quadratic manner at the $\Gamma$ point. This band touching is protected by symmetry: the two states at $\Gamma$ belong to the two-dimensional irreducible $E_{u}$ representation for the $D_{3d}$ point group. These features as well as the metallic phase are preserved under a compressive biaxial strain, as shown in Fig.~\ref{figQPT}(a).

Interesting scenario happens when a tensile biaxial strain is applied. As shown in Fig.~\ref{figQPT}(b), the Mexican-hat-like band gradually changes to a parabolic shape with increasing strain, and the system becomes a semimetal with a single Fermi point for a range of strain between $\sim 4\%$ and $\sim 8\%$. Further increasing strain transforms the system into a semiconductor [see Fig.~\ref{figQPT}(d)]. These results are consistent with the findings by Meng \emph{et al.}~\cite{meng2017metal} A phase diagram with the biaxial strain as the control parameter is illustrated in Fig.~\ref{figQPT}(e).

Our focus in the current work is the various emergent fermions that appear during the phase transitions. In the following, we analyze them one by one.

\subsection{Tilted anisotropic Dirac fermion}

\begin{figure}[t!]
\includegraphics[width=8cm]{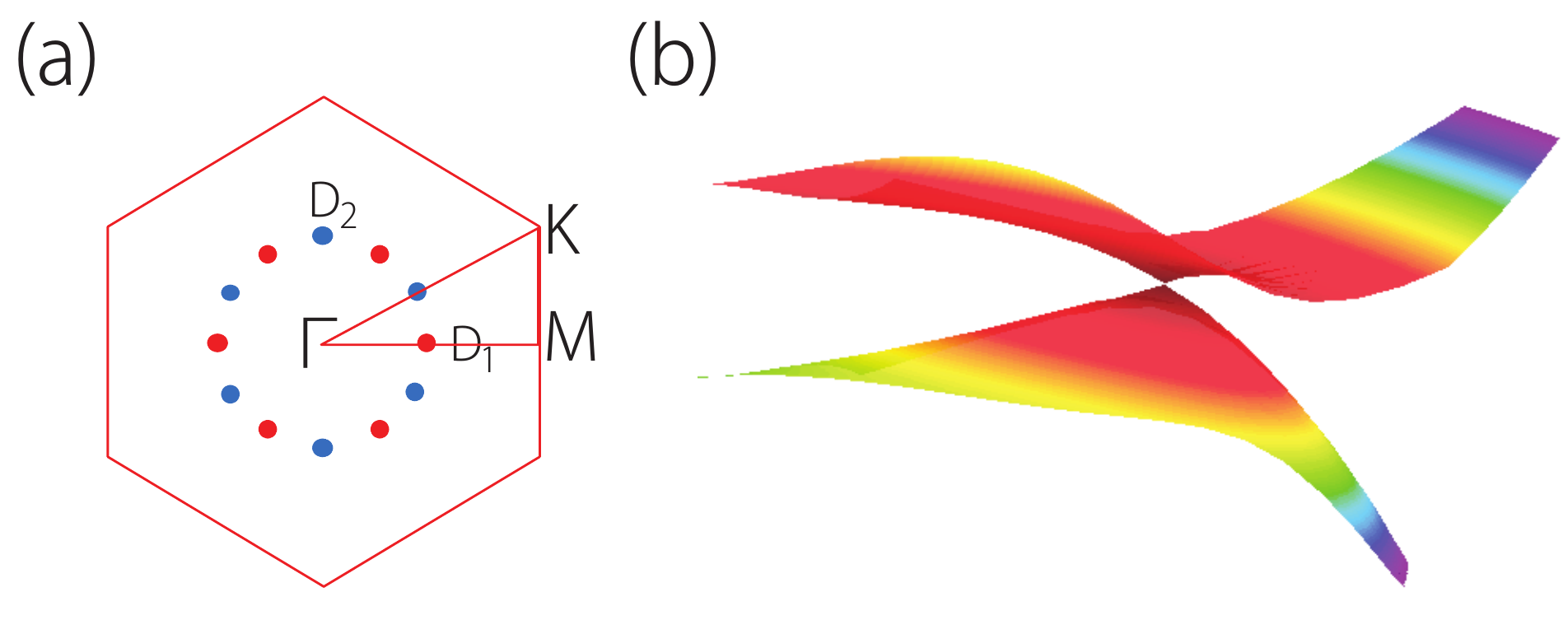}
\caption{2D tilted anisotropic Dirac points. (a) There are totally 12 Dirac points in the BZ. The red (blue) dots denote Dirac points $D_1$ ($D_2$). (b) Band dispersion around $D_1$. The dispersion around $D_2$ is similar, hence is not shown.}
\label{figDirac}
\end{figure}

In the metallic phase, there are two low-energy bands crossing the Fermi level. As we noted earlier, the two bands touch quadratically at the $\Gamma$ point, which is protected by the $D_{3d}$ point group symmetry. Furthermore, because one of the bands takes a Mexican-hat-like shape, the two bands linearly cross along the high-symmetry paths $\Gamma$-M and $\Gamma$-K, forming 2D Dirac points [see Fig.~\ref{figBand}(a)]. Due to the three-fold rotation and inversion symmetries, there are totally 12 Dirac points in the BZ [see Fig.~\ref{figDirac}(a)].

These Dirac points are protected by symmetry. For example, the Dirac point $D_1$ on $\Gamma$-M is protected by the vertical mirror plane $\mathcal{M}_{xz}$: the two bands along this path have opposite $\mathcal{M}_{xz}$ eigenvalues, hence their crossing point is protected.
Similarly, the Dirac point $D_2$ on $\Gamma$-K is protected by the two-fold rotation $\mathcal{C}_{2y}$ on this path. In addition, the system preserves the inversion symmetry $\mathcal{P}$ and time reversal symmetry $\mathcal{T}$. In the absence of SOC, the presence of $\mathcal{PT}$ symmetry guarantees that the Berry phase along any closed path must be quantized in unit of $\pi$. The Berry phase
\begin{equation}
\gamma_\ell=\sum_{n\in \text{occ.}}\oint_\ell \langle u_n(\bm k)|i\nabla_{\bm k}u_n(\bm k)\rangle \cdot d\bm k
\end{equation}
is defined for a locally gapped spectrum along a closed path $\ell$, $|u_n(\bm k)\rangle$ is the cell-periodic part of the Bloch eigenstate, and the band index $n$ is summed over the occupied bands below the local gap. Here, $\gamma_\ell$ for a closed path $\ell$ encircling the Dirac point must be $\pm \pi$, hence protecting the point against weak perturbations from opening a gap. We have numerically checked that such Berry phase is nontrivial in our DFT calculation. Thus, we see that the Dirac points here enjoy multiple protections by symmetries.

The band dispersion around a single Dirac point ($D_1$) is shown in Fig.~\ref{figDirac}(b), which indeed shows a Dirac-type linear dispersion. One also notes that the dispersion is anisotropic and the Dirac cone is tilted, in contrast with the Dirac dispersion in graphene which is isotropic and has a up-right Dirac cone. This is because that the Dirac points in graphene is located at the high-symmetry points K and K' which possess the three-fold rotational symmetry, whereas the Dirac points $D_1$ and $D_2$ here are located on the high-symmetry lines with reduced symmetries. Such kind of anisotropic Dirac points also appear in graphyne~\cite{malko2012competition}, group-V monolayers~\cite{lu2016multiple}, and Germanene on Al(111)~\cite{liu2015multiple}.

To characterize low-energy bands and the emergent fermions, we construct low-energy effective $k\cdot p$ models. Since the lower valence band (indicated by the arrow in Fig.~\ref{figQPT}) also participates in the metal-semimetal-semiconductor phase transition, to facilitate later discussion, we also include this band in the modeling. We perform the expansion around the $\Gamma$ point in the basis of the three states at $\Gamma$: the two degenerate states
corresponding to the $E_{u}$ irreducible representation and the lower nondegenerate state corresponding to the $A_{1g}$ representation. The model should respect the following symmetry operations: $\mathcal{C}_{3z}$, $\mathcal{M}_{xz}$, $\mathcal{P}$, and $\mathcal{T}$. Expanded up to $k$-quadratic order, the obtained effective Hamiltonian takes the form of
\begin{equation}\label{H0}
\begin{split}
    &H_{0}(\bm k)= \\
    &\left[\begin{array}{ccc}
    M_{1}+B_3 k^{2} & iAk_{x} & -iAk_{y} \\
    -iAk_{x}& M_{2}+B_1k^{2}_{x}+B_2k^{2}_{y}& (B_2-B_1)k_{x}k_{y} \\
    iAk_{y}& (B_2-B_1)k_{x}k_{y} & M_{2}+B_2 k^{2}_{x}+B_1 k^{2}_{y}
    \end{array}\right],
\end{split}
\end{equation}
where $k=|\bm k|$, and $A$, $B_1$, $B_2$, $B_3$, $M_1$, and $M_2$ are real parameters that may be obtained from fitting the DFT band structure (see Supplemental Material~\cite{supp}).

At the $\Gamma$ point, $k_x=k_y=0$, the energy of the three states are
\begin{equation}
E_{A_{1g}}=M_1, \qquad E_{E_u}=M_2.
\end{equation}
In the metallic as well as the semimetal phase, the $E_u$ doublet is above the $A_{1g}$ state, so we have $M_2>M_1$. However, after transition into the semiconductor phase, the $A_{1g}$ state becomes above the $E_u$ states, hence we have $M_2<M_1$.

Now we focus on the Dirac points in the metallic phase. We can project the 3-band model in Eq.~(\ref{H0}) onto the upper two ($E_u$) bands, and then expand around the Dirac points $D_1$ or $D_2$. Via straightforward calculations, we obtain the effective model for the Dirac fermions around $D_1$ [located at $\bm k=(k_D,0)$], given by
\begin{equation}\label{HD1}
\mathcal{H}_{D_1}(\bm q)=wq_x+v_x q_x\sigma_x+v_y q_y\sigma_y,
\end{equation}
where the wave vector $\bm q$ is measured from $D_1$, the $\sigma$'s are the Pauli matrices. The real parameters $w$, $v_x$, and $v_y$ here can be expressed in terms of the parameters in Eq.~(\ref{H0}), namely
\begin{equation}
w=\Lambda+B_2 k_D,\qquad v_x=\Lambda-B_2 k_D,
\end{equation}
\begin{equation}
v_y=\frac{(B_1-B_2)k_D+A(\zeta+\sqrt{1+\zeta^2})}{[1+(\zeta+\sqrt{1+\zeta^2})^2]^{1/2}},
\end{equation}
where
\begin{equation}
k_D=\sqrt{\frac{A^2-(M_1-M_2)(B_1-B_2)}{(B_3-B_2)(B_1-B_2)}},
\end{equation}
$\Lambda=\frac{1}{2\sqrt{1+\zeta^2}}[A+B_1(\sqrt{1+\zeta^2}-\zeta)k_D+B_3(\sqrt{1+\zeta^2}+\zeta)k_D]$, and $\zeta=[M_1-M_2+(B_3-B_1)k_D^2]/(2Ak_D)$. In fact, the form of the model (\ref{HD1}) can also be directly obtained from the symmetry requirement at the $D_1$ point, corresponding to the $C_{s}$ point group symmetry. Evidently, (\ref{HD1}) represents a tilted anisotropic Dirac fermion model, in which the first term $wq_x$ tilts the spectrum along the $x$ direction, and $v_x\neq v_y$ indicating the anisotropy.

In a similar way, the effective model around the Dirac point $D_2$ can be obtain, which takes the form of
\begin{equation}\label{HD2}
\mathcal{H}_{D_2}(\bm q)=wq_y+v_x q_y\sigma_x+v_y q_x\sigma_y.
\end{equation}
Here, $\bm q$ is measured from $D_2$, and the parameters $w$, $v_x$, and $v_y$ are the same as that in Eq.~(\ref{HD1}).

\subsection{2D double Weyl fermion}

For usual metal-semiconductor phase transitions, the semimetal state only appears as a critical state at the transition point. In sharp contrast, for the current system, semimetal occupies an extended range in the phase diagram [see Fig.~\ref{figQPT}(e)]. As we discuss below, this is because the Fermi point here is protected by symmetry. Hence, the semimetal state here may be more appropriately termed as a symmetry-protected semimetal.

\begin{figure}[t!]
\includegraphics[width=8cm]{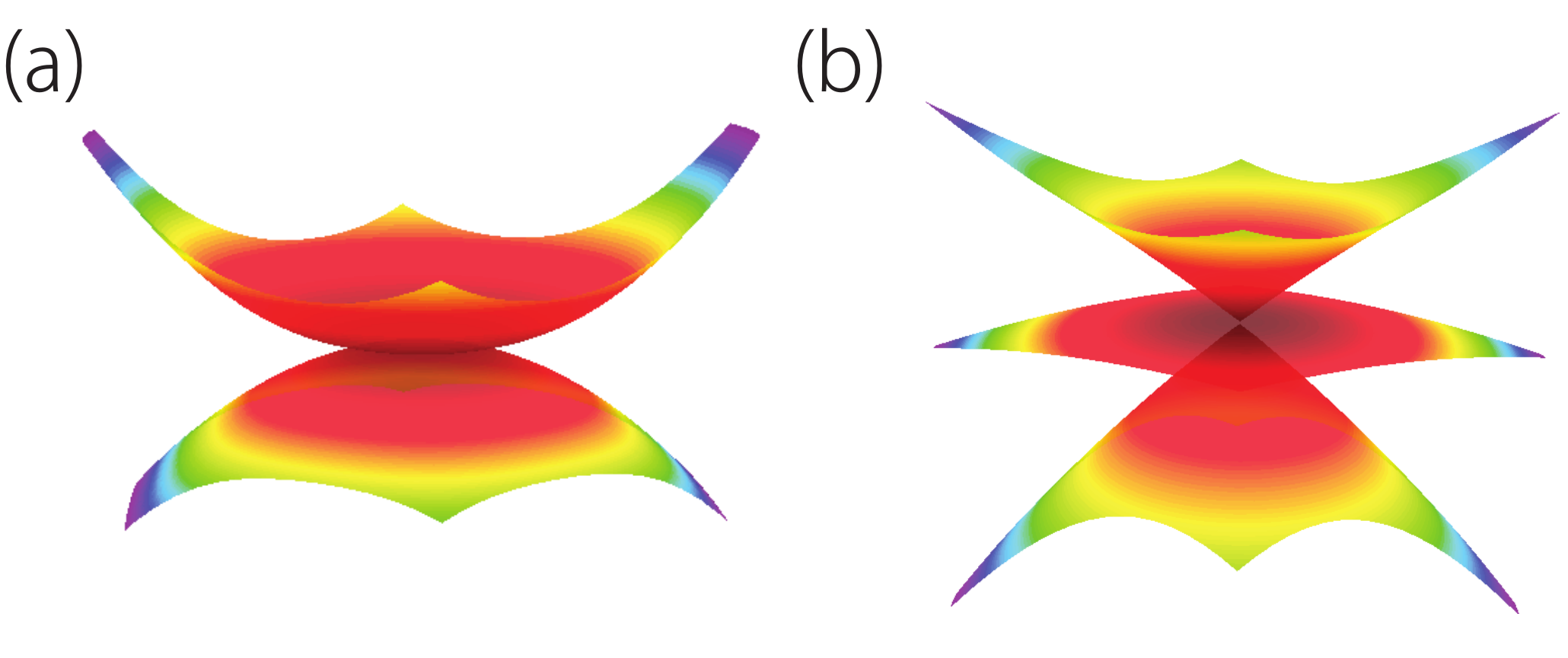}
\caption{Energy dispersion (around the band crossing point) correspond to (a) 2D double Weyl fermions, and (b) 2D pseudospin-1 fermions.}
\label{figDWTP}
\end{figure}

In the semimetal phase, the Fermi level lies exactly at the double Weyl point due to band filling [see Fig.~\ref{figQPT}(b)]. The 2D double Weyl point is formed by the quadratic touching between the conduction and valence bands [see Fig.~\ref{figDWTP}(a)], and is protected by symmetry because it corresponds to the $E_u$ two-dimensional irreducible representation for the $D_{3d}$ point group.

Again, we can derive an effective model to characterize the double Weyl fermions around this point. We may derive it from model (\ref{H0}), or, alternatively, directly obtain it using the symmetry constraints on the $E_u$ basis functions. Both approaches lead to the same result, given by (up to $k$-quadratic order)
\begin{equation}\label{HDW}
\mathcal{H}_\text{DW}(\bm k)=\alpha k^2+\beta \left[\begin{array}{cc}
                0 & k_-^{2}  \\
                k_+^{2} & 0
              \end{array}\right],
\end{equation}
where $k_\pm=k_x\pm ik_y$, and the parameters $\alpha$ and $\beta$ can be connected with the parameters in model (\ref{H0}) as $\alpha=(B_1+B_2)/2$ and
$\beta=(B_1-B_2)/2$.

The 2D double Weyl point features a Berry phase of $2\pi$. It has been previously discussed in the blue phosphorene oxide~\cite{zhu2016blue}. And without the diagonal term (first term in (\ref{HDW}) which breaks the particle-hole symmetry), the model also describes the $AB$-stacked bilayer graphene~\cite{mccann2006landau}. However, for bilayer graphene, there are two such points at K and K' points of the BZ related by time reversal symmetry. In comparison, for the Mg$_2$C monolayer, there is only a single 2D double Weyl point located at the $\Gamma$ point.

Due to its $2\pi$ Berry phase, the double Weyl fermion will exhibit weak localization at low-temperature, rather than weak anti-localization for the Dirac fermion~\cite{neto2009}. And due to its gapless quadratic dispersion, it will show a universal optical absorbance of $\pi \alpha \simeq 2.3\%$ for low-frequency electromagnetic waves~\cite{zhu2016blue}. Here, $\alpha=e^2/(\hbar c)$ is the fine structure constant. The frequency window for the universal optical absorbance is limited by the energy difference between the $E_u$ and $A_{1g}$ states, i.e., $|M_1-M_2|$ [see Eq.~(\ref{H0})], because beyond this value, the interband transition between the $A_{1g}$ band and the upper $E_u$ band would also contribute.

\subsection{2D pseudospin-1 fermion}

During the semimetal-semiconductor transition, the ordering between the $E_u$ doublet and the $A_{1g}$ state are switched: in the semimetal phase, $E_u$ is higher in energy, whereas in the semiconductor phase, $A_{1g}$ is higher. At the critical point for the phase transition, $E_u$ and $A_{1g}$ states overlap in energy and the gap closes at the $\Gamma$ point. Note that this overlap is also a consequence of symmetry---the states belong to different irreducible representations, otherwise they would repel each other and cannot overlap. At this critical point, the Fermi level cut through a single Fermi point where three bands meet each other [see Fig.~\ref{figDWTP}(b)], such that the low-energy fermions have an intrinsically three-component form.

To obtain an effective model describing the low-energy fermions for the critical point, we set $M_1=M_2=0$ in Eq.~(\ref{H0}), and close to the $\Gamma$ point, we may keep only the $k$-linear terms. The resulting model takes the simple form of
\begin{equation}\label{PS1}
\mathcal{H}_\text{PS-1}(\bm k)=A\bm k\cdot \bm S,
\end{equation}
where
\[
S_{x}=\left[\begin{array}{ccc}
                0 & i & 0 \\
                -i & 0 & 0 \\
                0 & 0 & 0
              \end{array}\right],
\qquad
S_{y}=\left[\begin{array}{ccc}
                0 & 0 & -i \\
                0 & 0 & 0 \\
                i & 0 & 0
              \end{array}\right],
\]
\begin{equation}
S_{z}=\left[\begin{array}{ccc}
                0 & 0 & 0 \\
                0 & 0 & i \\
                0 & -i & 0
              \end{array}\right]
\end{equation}
are three of the eight Gell-Mann matrices corresponding to spin-1, satisfying the angular momentum algebra $[S_i, S_j]=i\epsilon_{ijk}S_k$. It should be noted that $S$ is an emergent pseduspin degree of freedom, not the real particle spin. The fermion described by Eq.~(\ref{PS1}) is helical, with a well-defined helicity of $\pm 1$ and $0$ corresponding to the eigenvalues of the helicity operator $\bm k\cdot \bm S/k$. The $\pm 1$ branches are massless, whereas the $0$ branch has a flat dispersion.

Such pseudospin-1 particles have been studied in the cold-atom systems and photonic crystals~\cite{shen2010single,urban2011barrier,goldman2011topological,paavilainen2016coexisting}. Several extraordinary properties have been predicted for them. For example, they exhibit super Klein tunneling effect~\cite{shen2010single}, which means that the particle can penetrate an energy barrier with nearly perfect transmission for a wide range of incident angles. In addition, such particles also exhibit supercollimation effect~\cite{fang2016klein}, namely, they show guided unidirectional transport in the presence of a periodic potential, regardless of its initial direction of motion. The Mg$_2$C monolayer hence offers a realistic material platform to probe these fascinating physics.

\subsection{Effect of uniaxial strain}

\begin{figure}[t!]
\includegraphics[width=8cm]{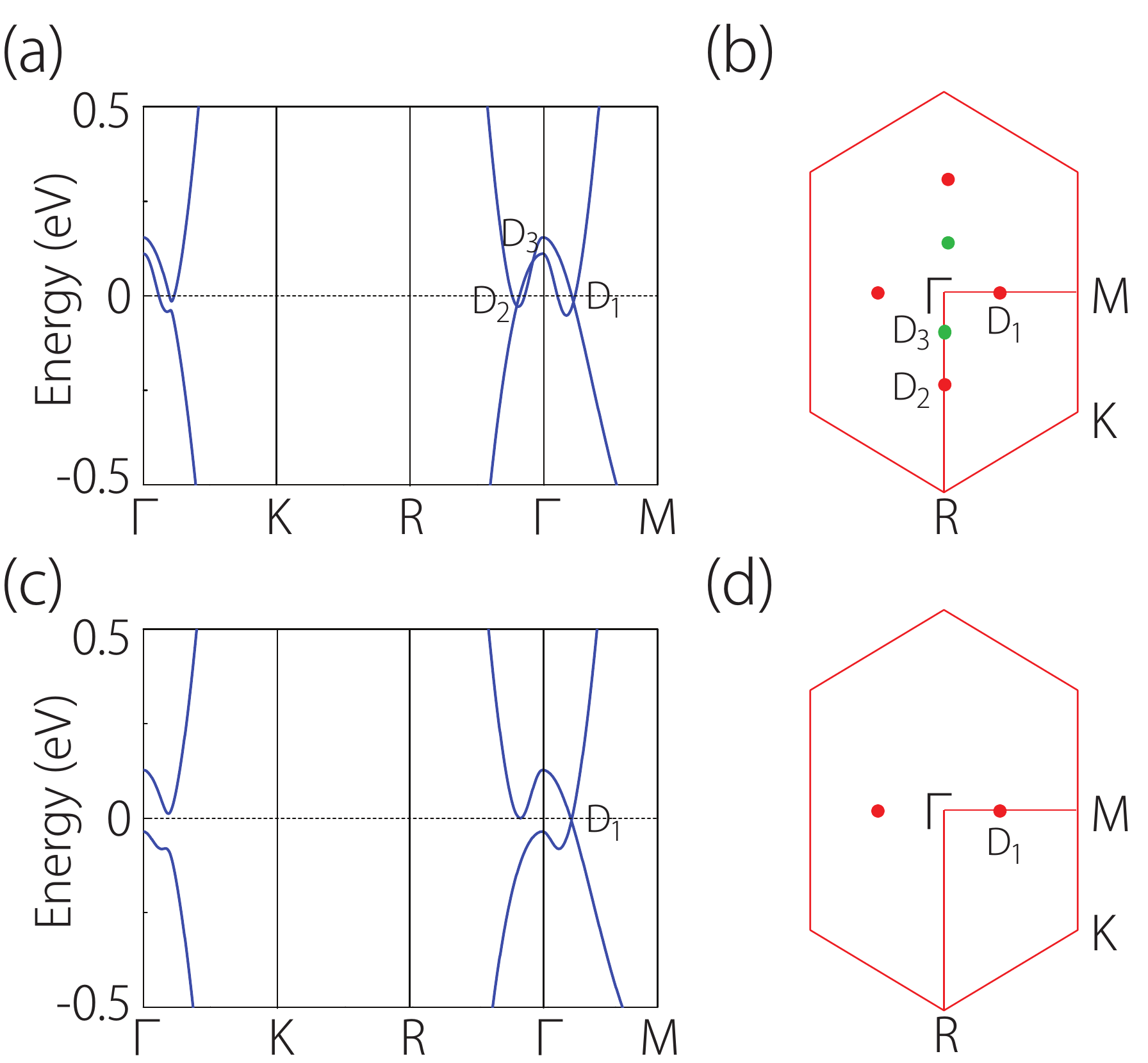}
\caption{(a) Electronic band structures of Mg$_{2}$C under $+1\%$ uniaxial strain in the armchair direction. $D_1$, $D_2$, and $D_3$ label the three Dirac points formed in the band structure. $D_3$ is a 2D type-II Dirac point. (b) Schematic figure showing the location of the Dirac points in the BZ. (c) and (d) are the corresponding figures for $+5\%$ uniaxial strain in the armchair direction. }
\label{figUniS}
\end{figure}

Unlike the biaxial strain, the uniaxial strain breaks the three-fold rotational symmetry and lowers the symmetry of the system, hence it is expected to transform the band structure in a qualitatively different way.

In Figure~\ref{figUniS}, we show the DFT results for uniaxial strains along the armchair direction. One observes the following features. First, due to the broken $\mathcal{C}_{3z}$, the point group symmetry reduces from $D_{3d}$ to $C_{2h}$, and the $E_u$ representation split into two one-dimensional representations $A_u$ and $B_u$. Hence, the original quadratic band touching point at $\Gamma$ no longer exists, and the corresponding degeneracy is lifted, as observed in Fig.~\ref{figUniS}(a,c). Second, the Dirac-type band crossing can still be observed along the high-symmetry paths. For small uniaxial strain, Dirac points appear on both $\Gamma$-M and $\Gamma$-R paths [labeled as $D_1$ and $D_2$ in Fig.~\ref{figUniS}(a)]. Their symmetry protections are the same as analyzed before, because the symmetries $\mathcal{M}_{xz}$, $\mathcal{C}_{2y}$, and $\mathcal{P}$ are preserved under this uniaxial strain. And because of the same symmetry constraints, the effective models for fermionic excitations around $D_1$ and $D_2$ take the same forms as Eq.~(\ref{HD1}) and Eq.~(\ref{HD2}).

One also notes that, at small strain, besides $D_1$ and $D_2$, there is another Dirac point $D_3$ on $\Gamma$-R, as indicated in Fig.~\ref{figUniS}(a). Interestingly, $D_3$ is a type-II Dirac point~\cite{xu2015structured,soluyanov2015aa,chang2017type} because the two crossing bands have the same sign for their slopes around $D_3$. The effective model for $D_3$ has the same form as Eq.~(\ref{HD2}), however, being a type-II point means that the tilt term dominates the spectrum, namely, we should have $|w|>|v_x|$.

At larger uniaxial strain ($>1\%$), we observe that the two bands are pulled apart on $\Gamma$-R, so the Dirac points $D_2$ is removed [see Fig.~\ref{figUniS}(c)]. However, the point $D_1$ on $\Gamma$-M is maintained. Hence for a range of strain, there are only two Dirac points in the BZ, as illustrated in Fig.~\ref{figUniS}(d).
Thus, the number of Dirac points can be effectively tuned by the uniaxial strain. The result for uniaxial strain along the zigzag direction is very similar, and hence is not shown.

\section{Discussion and Conclusion}

In this work, we reveal several unusual properties for the 2D material Mg$_2$C monolayer. We point out that the material possesses an intrinsic negative Poisson's ratio. This property is rare in nature. Recent studies on 2D materials have predicted several candidates that have negative Poisson's ratio, such as the black phosphorene~\cite{jiang2014negative}, $\delta$-phosphorene~\cite{wang2017delta}, borophene~\cite{zhou2014semimetallic}, penta-graphene~\cite{zhang2015penta}, Be$_5$C$_2$~\cite{wang2016semi}, and Zn$_2$C monolayer~\cite{meng2018unique}. In practice, this unusual property typically endows the material with enhanced toughness and shear resistance, and can lead to improved sound and vibration absorption~\cite{jiang2014negative}.

Techniques for applying strain on 2D materials have been well developed in recent years. For example, strain can be applied by using a beam bending
apparatus~\cite{conley2013bandgap} or by using an atomic force microscope tip~\cite{lee2008measurement}. Strains above 15\% have been demonstrated in experiment~\cite{lee2008measurement,bertolazzi2011stretching}. In general, tensile strains are easier to apply than compressive strains, because 2D materials tend to buckle under compression.

In experiment, the strain-induced phase transitions can be probed by the scanning tunneling spectroscopy (STS), which detects the local DOS. The gap-opening across the semimetal-semiconductor phase transition should give pronounced signals in transport measurement as well as in optical response. Especially, the unique signatures for the emergent fermions, including the universal optical absorbance in the semimetal phase, and the super Klein tunneling and the supercollimation effect near the semimetal-semiconductor transition can be probed in experiment. In addition, the characters for different emergent fermions may also be reflected in the Landau level (LL) spectrum. For example, the Dirac dispersion possess a zeroth LL (one for each spin species)~\cite{neto2009}, the double Weyl has two~\cite{mccann2006landau}, and the pseudospin-1 dispersion has a highly-degenerate zeroth LL due to the flat zero-helicity band~\cite{zhong2017three,Bradlyn2016}. These features can be detected by the magnetoinfrared spectroscopy or by STS experiment.

It should be noted that the unconventional emergent fermions and their unusual properties are mostly determined by symmetry. Thus, they are quite robust. For example, the qualitative features are not sensitive to the first-principles calculation methods (see HSE06 results in the Supplemental Material~\cite{supp}).

Finally, we also examine the 2D materials Be$_{2}$C and Ca$_{2}$C with the same type of lattice structure by replacing Mg with elements from the same group (see Supplemental Material~\cite{supp}). Unlike Mg$_{2}$C, we find that for Be$_{2}$C, it is a large gap semiconductor with the two $E_{u}$ bands at the valence band top, while the $A_{1g}$ band is pushed away. The system remains a semiconductor under a compressive biaxial strain $\sim -5\%$. The Ca$_{2}$C monolayer is metallic, and the pseudospin-1 point can emerge under a moderate biaxial strain $\sim 5\%$.

In conclusion, we have discovered several fascinating properties for the $2$D hexacoordinated Mg$_{2}$C monolayer. We analyze the material's phonon spectrum and identify the Raman active modes. We show that it has excellent mechanical flexibility and an intrinsic in-plane negative Poisson's ratio. Most interestingly, its electronic structure can be effectively tuned by both biaxial and uniaxial strains, during which several types of unconventional fermionic excitations emerge at low energy. For biaxial strains, we find that during a metal-semimetal-semiconductor quantum phase transition, titled anisotropic Dirac fermions (around 12 Dirac points), $2$D double Weyl fermions, and 2D pseudospin-$1$ fermions can appear, which are tied with the symmetry of the system. For uniaxial strains, a reduced number of Dirac points can be preserved including 2D type-II Dirac points. Effective models are constructed to characterize their unique features. The $2$D hexacoordinated Mg$_{2}$C monolayer thus provides a promising platform to investigate the intriguing properties of unconventional types of fermions beyond the isotropic Dirac fermions in graphene. We hope that our theoretical work will facilitate the experimental studies on this new 2D material towards both fundamental discoveries and potential applications.

\begin{acknowledgements}
The authors thank S. Wu and D. L. Deng for valuable discussion. This work is supported by the Singapore Ministry of Education Academic Research Fund Tier 2 (MOE2015-T2-2-144). We acknowledge computational support from the Texas Advanced Computing Center and National Supercomputing Centre Singapore.

Shan-Shan Wang and Ying Liu contributed equally to this work.
\end{acknowledgements}


\bibliography{Mg2C_ref}


\end{document}